\documentclass{article}
\usepackage{colordvi,epsfig,bm,amssymb}
\newcommand{\bbbr}      {{\mathbb{R}}}                  
\newcommand{\bbbone}{{\mathbb{I}}}

\newcommand{\rd} {\mathrm d}

\newcommand{\ri} {\mathrm i}

\newcommand{\be}{\begin{equation}} 
\newcommand{\ee}{\end{equation}}  
\newcommand{\bea}{\begin{eqnarray}}
\newcommand{\eea}{\end{eqnarray}}
\begin{document}
\title{Disentangling Giant Component and Finite Cluster Contributions in Sparse Matrix Spectra}
\author{Reimer K\"uhn\\
Mathematics Department, King's College London, Strand, London WC2R 2LS,UK}
\maketitle
\begin{abstract}
\noindent
We describe a method for disentangling giant component and finite cluster contributions to sparse
random matrix spectra, using sparse symmetric random matrices defined on Erd\H{os}-Renyi graphs
as an example and test-bed.
\end{abstract}
\section{Introduction}
While there has been considerable recent progress in the understanding of sparse random matrix spectra
\cite{RodgBray88, BirMon99, SemerjCugl02, Dorog+03, Ku08, Rog+08, Erd+13,Bord+14}, there is at least one 
important open problem which has not yet been properly addressed, viz. the disentangling of contributions 
to limiting spectra coming from the giant component and from finite clusters, respectively. This is of 
particular relevance when separating pure point (localized) and absolutely continuous components of 
random matrix spectra: contributions from finite clusters are trivially localized, yet what we are 
mainly interested in is to identify pure point contributions to random matrix spectra originating from 
the giant component, rather than the `trivial contaminations' of these coming from finite clusters.

The present note is meant to address and solve this very problem. The solution is relevant also to the
analysis of other forms of collective phenomena on networked systems, such as the analysis of infection
dynamics or of network models of systemic risk in finance. We describe our method for the spectral problem 
of weighted adjacency matrices. The same method can be used to evaluate spectra of (weighted) graph 
Laplacians \cite{Ku08}, sparse Markov Matrices \cite{Ku15,Ku15b}, or non-Hermitian sparse matrices 
\cite{RogPer09}.

\section{Spectral Density and Resolvent}
We are interested in evaluating the spectral density of sparse matrices $A$ of the form
\be
A_{ij}= c_{ij} K_{ij}\ ,
\label{defA}
\ee
in which $C=(c_{ij})$ is a sparse connectivity or adjacency matrix describing a finitely coordinated 
random graph, and $K=(K_{ij})$ a matrix of edge weights. We take both $C$ and $K$ to be real symmetric
matrices.

The spectral density of $A$ is obtained from the resolvent using the Edwards Jones approach 
\cite{EdwJon76} as
\be
\rho_A(\lambda) = \frac{1}{\pi N}~\lim_{\varepsilon\to 0} {\rm Im~Tr~} (\lambda_{\varepsilon} \bbbone - A)^{-1}
= -\frac{2}{\pi N}~\lim_{\varepsilon\to 0} {\rm Im}~\frac{\partial}{\partial \lambda} \ln Z_N\ ,
\label{rho-res}
\ee
with $\lambda_{\varepsilon}= \lambda-\ri\varepsilon$ and 
\be
Z_N = \int \prod_{i=1}^N \frac{\rd u_i}{\sqrt{2\pi/\ri}}~\exp\Big\{
-\frac{\ri}{2}\sum_{i,j}  \big(\lambda_{\varepsilon} \delta_{ij} - 
A_{ij}\big)\ u_i u_j\Big\}\ .
\label{ZN}
\ee
This gives
\be
\rho_A(\lambda) = \frac{1}{\pi N}~\mbox{Re}~\sum_i \langle u_i^2\rangle\ ,
\ee
where $\langle \dots \rangle$ is an average w.r.t. the complex Gaussian
measure defined by (\ref{ZN}). Only single-site variances are needed for
the evaluation. The role of $\varepsilon$ in these equations is to ensure that
integrals converge even for $\lambda$ in the spectrum of $A$, and the limit
$\varepsilon\to 0$ should be taken at the end of the calculation. However, as 
demonstrated elsewhere \cite{Ku08}, there is a second role of $\varepsilon$, 
namely as a regularizer of the spectral density, and a small non-zero value
of $\varepsilon$ must be kept in the evaluation of spectral densities in order
to expose pure point contributions to spectra.

This representation can be used to evaluate spectral densities for large single 
problem instances in terms of cavity recursions \cite{Rog+08} as detailed below.
In the thermodynamic limit these can be interpreted as stochastic recursions,
giving rise to a self-consistency equation for pdfs of (inverse) variances 
of cavity marginals. Alternatively, thermodynamic limit results are obtained by 
averaging (\ref{rho-res}) over the ensemble of random matrices considered, using 
replica or to perform the average.
\section{Cavity Analysis}
\label{sec:cav}
As demonstrated in \cite{Rog+08}, one can use the cavity method to evaluate the marginals 
of the complex Gaussian defined by (\ref{ZN}) which are needed in the evaluation of (\ref{rhoN}). 
We briefly repeat the reasoning here, both for completeness and in order to prepare a generalization
that keeps track of the information whether a site belongs to the giant component or to one of
the finite clusters of the system.

For a single-site marginal we have the representation
\be
P(u_i) \propto \exp\Big\{-\frac{\ri}{2}\lambda_{\varepsilon}\, u_i^2\Big\}
\int \prod_{j\in \partial i} \rd u_j \, \exp\Bigg\{\ri \sum_{j\in \partial i}
K_{ij} u_i u_j\Bigg\} P_j^{(i)}(u_j)\ ,
\label{ss-marg}
\ee
with $\partial i$ denoting the set of sites connected to $i$ (which may or may not 
be empty) and $P_j^{(i)}(u_j)$ denoting the complex cavity weight of $u_j$. On
a (locally) tree-like graph one may write down a recursion for the cavity weights,
\be
P_j^{(i)}(u_j) \propto \exp\Big\{-\frac{\ri}{2}\lambda_{\varepsilon}\, 
u_j^2\Big\} \prod_{\ell\in \partial j\setminus i} \int \rd u_\ell \, 
\exp\Big\{\ri K_{j\ell} u_j u_\ell\Big\} P_\ell^{(j)}(u_\ell)\ .
\label{ss-cmarg}
\ee
As demonstrated in \cite{Rog+08}, recursions of this type are self-consistently 
solved by complex Gaussians of the form
\be
P_j^{(i)}(u_j) = \sqrt{\frac{~\omega_j^{(i)}}{2\pi}} \, 
\exp\Big\{-\frac{1}{2}\omega_j^{(i)} u_j^2\Big\}\ ,
\label{cav-m}
\ee
which transforms Eq. (\ref{ss-cmarg}) into a recursion for the $\omega_j^{(i)}$,
\be
\omega_j^{(i)} = \ri \lambda_{\varepsilon} +
\sum_{\ell\in \partial j\setminus i} \frac{K_{j\ell}^2}{\omega_\ell^{(j)}}\ .
\label{cav-rec}
\ee
This recursion can be solved iteratively for large single instances. 

In terms of the solution, the spectral density is given by
\be
\rho_A(\lambda) =  \frac{1}{\pi N}\,\mbox{Re}\,
\sum_{i} \,\langle u_i^2\rangle\ ,
\label{rhoN}
\ee
with
\be
\langle u_i^2\rangle = \frac{1}{\omega_i}
\ee
and
\be
\omega_i = \ri \lambda_{\varepsilon}\, +
\sum_{j\in \partial i} \frac{K_{ij}^2}{\omega_j^{(i)}}
\label{ss-var}
\ee

Alternatively in the infinite system limit of a random system one can interpret 
Eq. (\ref{cav-rec}) as a stochastic recursion for the collection $\{\omega_j^{(i)}\}$ of 
random inverse cavity variances,  which in turn generates a recursion for the pdf
$\pi(\omega)$ of the $\omega_j^{(i)}$. 

\subsection{Averaging Stochastic Recursions}
\label{sec:av}
Averaging single instance cavity equations to obtain equations for distributions $\pi(\omega)$
of inverse cavity variances valid for the thermodynamic limit follows standard reasoning. We
have
\be
\pi(\omega) = \sum_{k\ge 1} p(k) \frac{k}{c} \int \prod_{\nu=1}^{k-1}
\rd \pi(\omega_\nu)\, \Big\langle\delta(\omega- \Omega_{k-1})\Big\rangle_{\{K_\nu\}}
\label{piom}
\ee
where $p(k) \frac{k}{c}$ is the probability to be connected to a site of degree $k$, and 
\be 
\Omega_{k-1}= \Omega_{k-1}(\{\omega_\nu,K_\nu\})=
\ri \lambda_{\varepsilon}+ \sum_{\nu=1}^{k-1} \frac{K_\nu^2}{\omega_\nu}\ .
\label{Om}
\ee
We use $\langle \dots \rangle_{\{K_\nu\}}$ to denote an average over the set of (independent) edge weights
appearing in the argument.

Similarly, the spectral density in the thermodynamic limit is given by
\be
\rho(\lambda) = \frac{1}{\pi}\,\mbox{Re}\, \,\sum_{k\ge 0} 
p(k) \int \prod_{\nu=1}^{k} \rd \pi(\omega_\nu)\,\ \frac{1}{\Omega_k(\{\omega_\nu,K_\nu\})}
\label{eq:StDos}
\ee

The problem with this approach is that it does {\em not\/} separate contributions to the limiting 
spectral density coming from the giant component and from finite clusters which are also represented 
in the graph-ensemble. 

For large single instances, one could of course always identify the largest component of a system,
restrict the cavity analysis of spectra to that largest component and subsequently average it over 
many realizations to obtain ensemble averages (albeit only finite-size approximations thereof).

In what follows we shall revisit the cavity analysis of sparse matrix spectra, and combine it with
a corresponding cavity analysis of the percolation problem on the graph for which random matrix spectra
are being evaluated, so as to disentangle giant component and finite cluster contributions to limiting
spectral densities.
\section{Cavity Approach and Ensemble Averaging Revisited}
In order to disentangle contributions to the spectral density coming from finite clusters and the 
giant component of a graph, respectively, we need to supplement the cavity analysis of Sect. 
\ref{sec:cav} by a component that allows one to keep track of the information whether a site belongs 
to the former or the latter.

To that end we use ideas developed for the analysis of the percolation problem of random graphs
\cite{New+01}. Rather than directly analysing percolation in terms of the fraction $p_{\rm gc}$ of 
vertices that belong to the giant cluster of a graph we use indicator-variables $n_i\in\{0,1\}$ 
signifying whether individual sites $i$ belong to the giant cluster of a graph ($n_i=1$) or whether, 
on the contrary, they belong to one of the finite clusters of the system ($n_i=0$).

For these we then have 
\be
n_i = 1 - \prod_{j\in\partial i}\big(1-n_j^{(i)}\big)
\label{ss-perc}
\ee
where $n_j^{(i)}$ is a cavity indicator variable signifying whether site $j$ does ($n_j^{(i)}=1$) or 
does not ($n_j^{(i)}=0$) belong to the giant cluster on the cavity graph, from which site $i$ and the edges 
connected to it have been removed. The cavity indicator variables then satisfy the recursion
\be
n_i^{(j)} = 1 - \prod_{\ell\in\partial j\setminus i}\big(1-n_\ell^{(j)}\big)\ .
\label{ss-cperc}
\ee
The structure of these equations for the indicator and the cavity indicator variables clearly mimics that
for the single-site marginals Eq. (\ref{ss-marg}) and the cavity marginals Eq. (\ref{ss-cmarg}), respectively.

In the large system limit of a random graph one can interpret Eq. (\ref{ss-cperc}) as a stochastic recursion
for the collection $\{n_i^{(j)}\}$ of cavity indicator variables that supplements the recursion Eq. (\ref{cav-rec})
for the inverse cavity variances  $\{\omega_i^{(j)}\}$. Combining the two then in turn generates a recursion for
the {\em joint distribution\/} $\pi(\omega,n)$ of inverse cavity variances {\em and\/} cavity indicator variables,
which take the form
\be
\pi(\omega,n) = \sum_{k\ge 1} p(k) \frac{k}{c} \sum_{\{n_\nu\}} \int \prod_{\nu=1}^{k-1}
\rd \pi(\omega_\nu,n_\nu) \, \Big\langle\delta(\omega- \Omega_{k-1})\Big\rangle_{\{K_\nu\}}\times
\delta_{n,1-\prod_{\nu=1}^{k-1}(1-n_\nu)}
\label{piomn}
\ee
From the solution of this equation one obtains the limiting spectral density as a sum of two contributions,
one of these ($\rho_{\rm gc}$) coming from the giant cluster, the other ($\rho_{\rm fc}$) from the collection
of finite clusters,
\be
\rho(\lambda)= \rho_{\rm gc}(\lambda) + \rho_{\rm fc}(\lambda)\ ,
\ee
with
\bea
\rho_{\rm gc}(\lambda) &=& \frac{1}{\pi}\,\mbox{Re}\, \,\sum_{k\ge 0} 
p(k) \sum_{\{n_\nu\}}\int \prod_{\nu=1}^{k} \rd \pi(\omega_\nu,n_\nu)\,\ \frac{1}{\Omega_k(\{\omega_\nu,K_\nu\})}
\times \delta_{1,1-\prod_{\nu=1}^{k}(1-n_\nu)}\ ,
\label{rhogc}\\
\rho_{\rm fc}(\lambda) &=& \frac{1}{\pi}\,\mbox{Re}\, \,\sum_{k\ge 0} 
p(k) \sum_{\{n_\nu\}}\int \prod_{\nu=1}^{k} \rd \pi(\omega_\nu,n_\nu)\,\ \frac{1}{\Omega_k(\{\omega_\nu,K_\nu\})}
\times \delta_{0,1-\prod_{\nu=1}^{k}(1-n_\nu)}\ ,
\label{rhofc}
\eea
with $\Omega_k(\{\omega_\nu,K_\nu\})$ as defined above.

Eq. (\ref{piomn}) is efficiently solved by a population dynamics algorithm \cite{MezPar01}, and the 
giant component and finite cluster contributions to the spectral density are evaluated by sampling 
from the equilibrium distribution of the population dynamics.

Both $\pi(\omega,1)$ and $\pi(\omega,0)$ have support in the complex half-plane Re $ \omega\ge 0$. 
As argued in \cite{Ku08}, a pure point contribution is signified by a singular component of $\pi(\omega,n)$ 
with support on the imaginary axis $\omega \in \ri\,\bbbr$.

\section{Results and Discussion}
In what follows, we briefly illustrate the workings of our method by providing sample spectra of
sparse matrices of the type (\ref{defA}). Here we present results for matrices defined on a sparse
Erd\H{o}s-Renyi graph of mean connectivity $c=2$. It goes without saying that other matrix and graph
ensembles can be analysed in the same way, in the sense that the method of disentangling giant and
finite cluster distributions described here is not restricted to Erd\H{o}s-Renyi graphs but works
for any system in the configuration model class, as well as for spectra of weighted graph Laplacians
\cite{Ku08} or of sparse random stochastic matrices \cite{Ku15, Ku15b}.

\begin{figure}[ht]
\begin{center}
\epsfig{file = 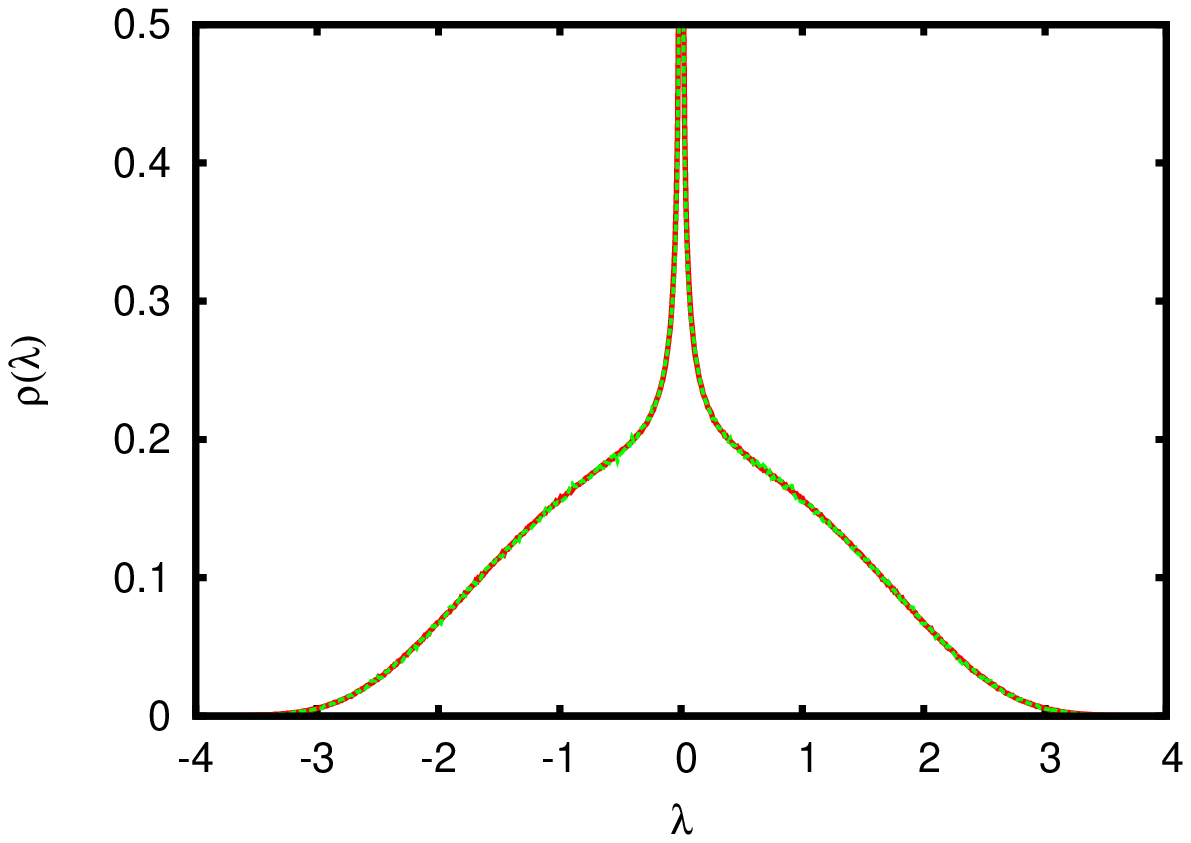, width=0.475\textwidth}\hfil
\epsfig{file = 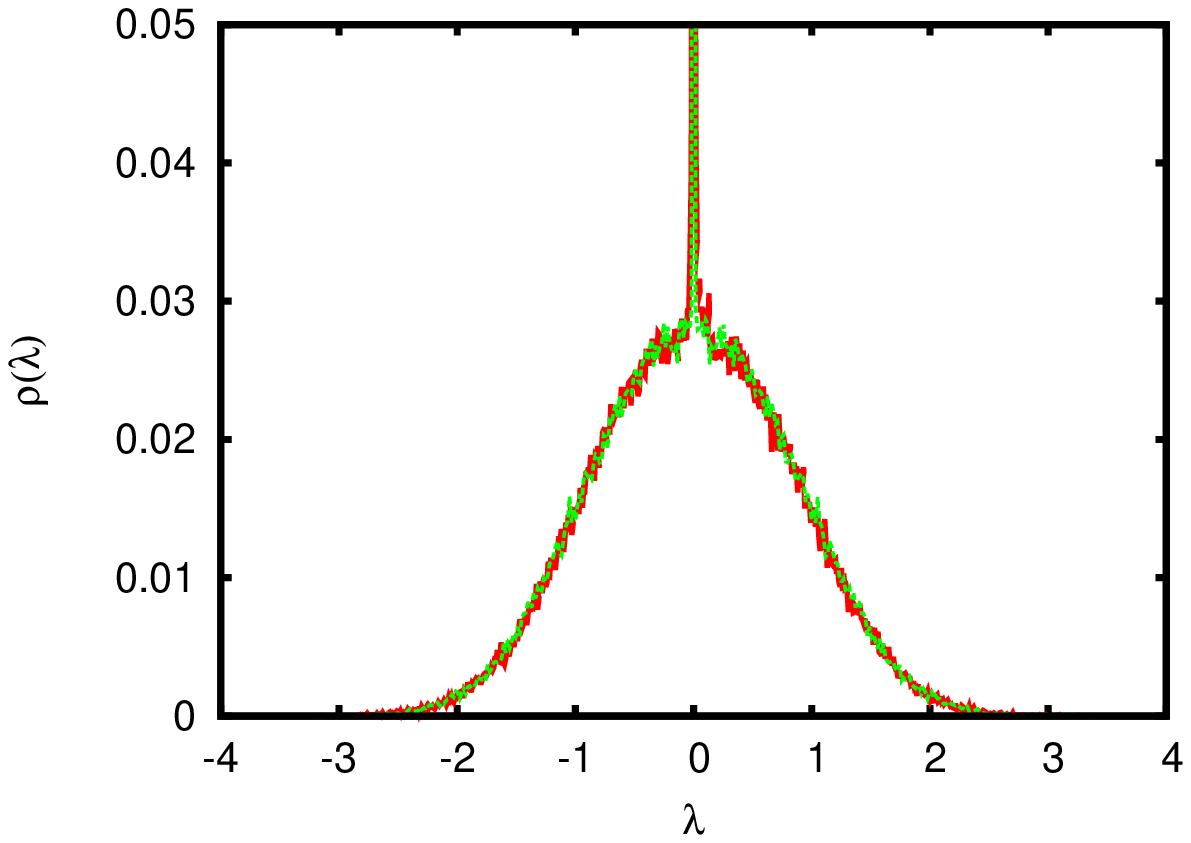, width=0.475\textwidth}\hfil
\end{center}
\caption{(Colour online) Spectral density of a random matrix defined on an Erd\H{o}s-Renyi random graph of 
mean connectivity $c=2$, with link weights normally distributed with standard deviation $\sigma=1/\sqrt c$. 
Left panel: giant-cluster contribution. Right panel: finite-cluster contribution. In both panels, full red 
lines represent results for the limiting spectral density obtained via population dynamics, while green 
dashed lines are simulation results using graphs of $N=500$ vertices, averaged over 5000 random instances.
Note the different vertical scale in the right panel.}
\end{figure}

In Fig. 1, we present the spectrum of a matrix with Gaussian random edge weights of standard deviation 
$\sigma=1/\sqrt c$ on the edges of the Erd\H{o}s-Renyi graph, separately exhibiting the contributions 
coming from the giant cluster and from the collection of finite clusters. The former occupies a fraction 
$p_{\rm gc}\simeq 0.796812$ of the entire system. We also compare our results with simulations, associating 
the giant cluster with the largest finite cluster of each realization of the system, and all other 
components with the collection of finite clusters, finding excellent agreement with theoretical results.
Note that the finite cluster results displayed in the right panel of Fig. 1 are slightly noisier than
those pertaining to the giant cluster, as a smaller fraction of updates in the population dynamics 
corresponds to finite cluster contributions.

\begin{figure}[ht]
\begin{center}
\epsfig{file = 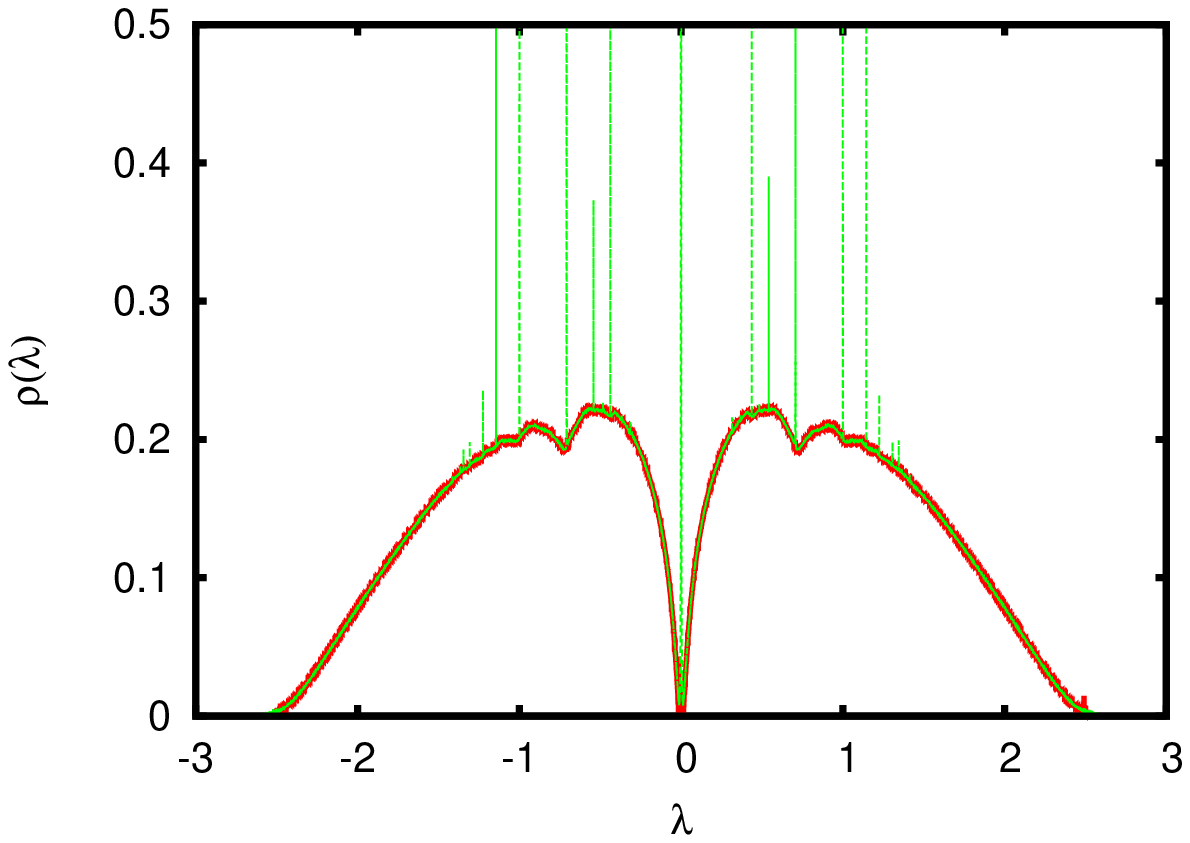, width=0.475\textwidth}\hfil
\epsfig{file = 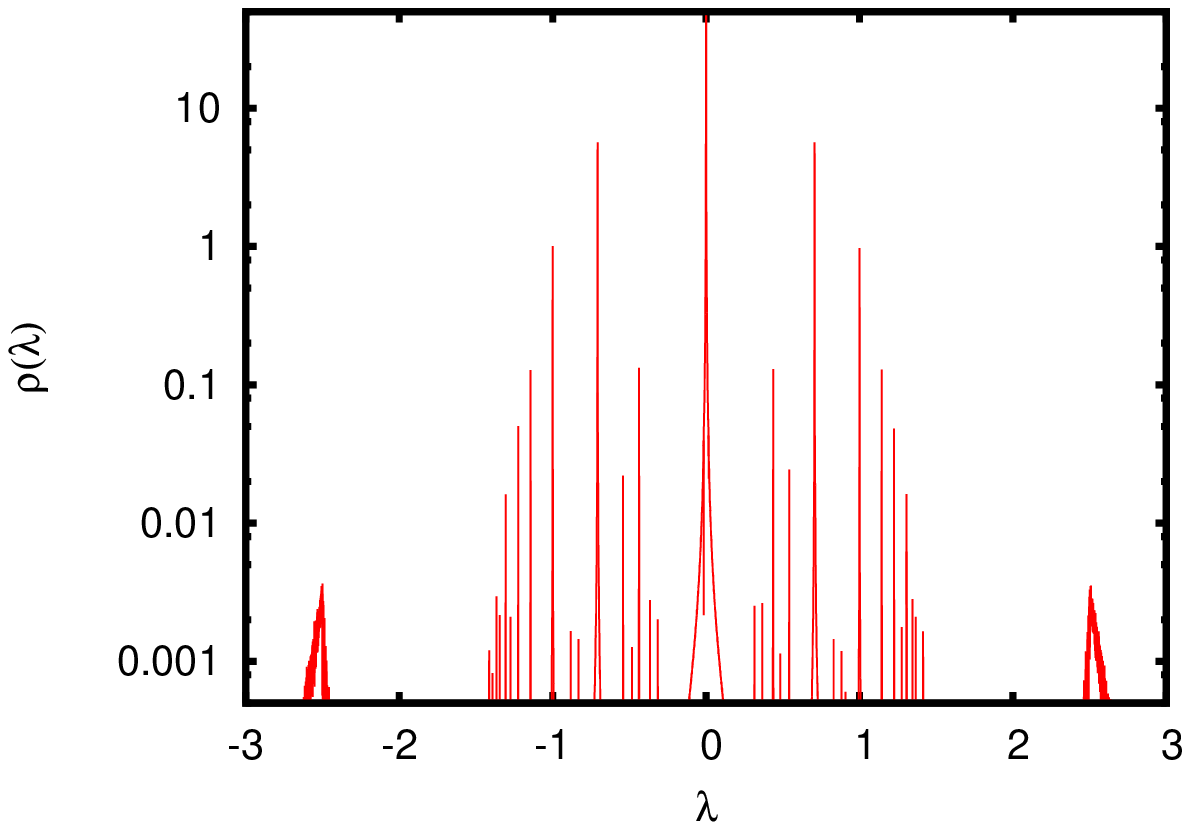, width=0.475\textwidth}\hfil
\end{center}
\caption{(Colour online) Spectral density of the adjacency matrix of the giant cluster of an 
Erd\H{o}s-Renyi random graph of mean connectivity $c=2$, with link weights chosen as $1/\sqrt c$. 
Left panel: continuous density of states (thick full red line) and total density of states including 
atoms (green dashed line). Right panel: spectrum of {\em localized\/} states on the giant cluster 
regularized at $\varepsilon = 10^{-3}$. The central part of the panel exhibits atoms in the bulk of
the spectrum, whereas the two bands in the vicinity of $\pm 2.5$ correspond to fully localized bands
of states, separated by mobility edges from the bulk of the spectrum. The normalization is chosen 
such that the total DOS integrates to the fraction of sites contained in the percolating cluster
(see Eq. (\ref{int-gcDOS})).}
\end{figure}

In Fig. 2 we show results for the spectrum of the adjacency matrix on the giant cluster of an 
Erd\H{os}-Renyi graph, with edge weight set at $1/\sqrt c$. As shown in \cite{BauGol01}, all 
eigenvalues which are eigenvalues of finite trees will also appear as eigenvalues of the adjacency 
matrix of the giant component of the system, and correspond to localized states. The left panel of Fig. 2
exhibits a few of these, namely the ones with the largest weights appearing in the giant component
spectrum; the weight of the remaining atoms is too small, entailing that these are `drowned' in the 
continuum at the resolution (and regularization) chosen in the figure. In the right panel the continuum
contribution is subtracted, so that it exhibits just the contribution of localized states to the 
spectrum of the giant component of the system, regularized at $\varepsilon= 10^{-3}$. Note that one 
effect of regularization is to broaden each $\delta$-peak into a Lorentzian of width $\varepsilon$, 
which is clearly visible for the peaks with the largest weight in the spectrum. The system also 
exhibits Anderson localization, entailing that {\em all\/} states with $|\lambda| > \lambda_c \simeq 
2.50$ are localized; this corresponds to the two bands of states at $|\lambda| > \lambda_c$ in the
right panel of Fig. 2. As expected we find {\em all\/} states on finite clusters to be localized.

A numerical integration (using a trapeze-rule) of the total density of states on the giant cluster gives
\be
\int \rd \lambda\, \rho_{\rm gc}(\lambda) \simeq 0.7969
\label{int-gcDOS}
\ee
which is agrees very well with the expected result $p_{\rm gc}\simeq 0.796812$, i.e. the fraction of 
vertices of the system in the giant cluster. Doing the integration for the (absolutely)-continuous 
component of the giant-cluster spectrum gives
\be
\int \rd \lambda\, \rho^{\rm (ac)}_{\rm gc}(\lambda) \simeq 0.7153
\ee
entailing that a fraction
\be
f_{\rm gc}^{\rm (loc)} \simeq \frac{0.7969-0.7153}{0.7969} \simeq 0.1020\ ,
\ee
i.e. approximately 10\% of all states on the giant cluster are localized.

To summarize, by combining approaches to percolation on random graphs and to the evaluation of sparse
matrix spectra we have presented a method that allows to separately evaluate contributions to sparse
matrix spectra coming from the giant cluster and from finite clusters, respectively. Our results are
confirmed to a high precision by numerical simulations, even at moderate system size. 

By further disentangling the absolutely continuous and pure point contribution to limiting spectra, we 
are able to give a precise estimate of the fraction of states  {\em on the giant cluster\/} that are 
localized. We are not aware of a previous such estimate, although a method to estimate the weight of 
the peak at $\lambda=0$ was recently devised by Bordenave et al. \cite{Bord+11}.

We expect our method to be useful for the analysis of other phenomena described in terms of networked 
systems, including e.g. the spread of diseases or computer viruses, the behaviour of random walks or 
the performance of search algorithms on networks.

A interesting field of research, where our results can provide a crucial ingredient of the analysis
is the investigation of localization phenomena, where it is important to avoid contamination of results 
from finite cluster contributions.

\paragraph{Acknowledgements} Illuminating discussions with Justin Salez and with Peter Sollich are 
gratefully acknowledged.

\bibliography{../../MyBib}
\end{document}